\documentclass[10pt,twocolumn]{article}
\usepackage[numbers,comma]{natbib}
\usepackage{graphicx}
\title{Resolution of Identity Crisis of Events in Pile-up}
\author{Avinash A. Deshpande \& Harsha Raichur \\Raman Research Institute, C. V. Raman Avenue, Sadashivanagar, Bangalore 560080, India} 
\date{} 
\begin{document}
\twocolumn[  
\begin{@twocolumnfalse} 
\maketitle
\begin{abstract}
Mutually uncorrelated random discrete events, manifesting 
a common basic process, are examined often in terms of 
their occurrence rate as a function of one or more of 
their distinguishing attributes, such as measurements 
of photon spectrum as a function of energy.
Such rate distributions obtained from the observed
attribute values for an ensemble of events will correspond
to the ``true" distribution only if the event occurrence  
were {\it mutually exclusive}. However, due to finite 
resolution in such measurements, the problem of event 
{\it pile-up} is not only unavoidable, but also increases 
with event rate. Although extensive simulations to estimate 
the distortion due to pile-up in the observed rate 
distribution are available, no restoration 
procedure has yet been suggested. Here we present an 
elegant analytical solution to recover the underlying 
{\it true} distribution. Our method, based on Poisson 
statistics and Fourier transforms, is shown to perform
as desired even when applied to distributions that are 
significantly distorted by pile-up. Our recipes for
correction, as well as for prediction, of pile-up are
expected to find ready applications in a wide 
variety of fields, ranging from high-energy physics
to medical clinical diagnostics,
and involving, but not 
limited to, measurements of count-rates and/or spectra 
of incident radiation using Charge Coupled Devices (CCDs) or other similar devices.\\
\end{abstract}
\end{@twocolumnfalse} 
] 

To formulate the {\it pile-up} problem analytically, 
let us denote the {\it true} and {\it apparent} 
event-rate distribution by $\lambda(S)$ and  $\lambda_a(S)$ 
respectively, where $S$ is a chosen attribute of the events 
we are interested in. Let $\Delta$ be the (spatio-temporal)
resolution with which these measurements are made.
The basic situation in which `pile-up' occurs
can be illustrated using the following simple example. 
Consider two events (say, with 
$S=S_1$ \& $S_2$) that occur within the measurement 
resolution $\Delta$. Since both the events occur within 
$\Delta$ they would be mistaken for a 
single event with an attribute $S_{sum}=S_1+S_2$. Therefore, 
instead of registering two events, one each at $S=S_1$ 
and $S=S_2$, only one event at $S=S_{sum}$ would be noted. 
As a result, the rate of events at $S_1$ and at $S_2$ 
are underestimated, with a corresponding overestimation 
of event rate at $S_{sum}$.
 
Generalizing the above example, if $n$ events (say at, $S_i$, $i=1 \to n$; $n>1$) 
were to overlap, then for each such occurrence, the event 
count at $S_{sum} = \sum_{i=1}^{n} S_i$ is not only wrongly incremented by one, 
but a count is missed at each $S_i$. Consequently, 
in addition to this mistaken identity in `$S$', even the net 
event count suffers a deficit of $n$-1, for each unresolved 
occurrence of $n$ events. Thus, in general, any measured $\lambda_a(S)$ deviates from the 
corresponding {\it true} distribution $\lambda(S)$ due to a finite 
probability of unresolved events occurring within $\Delta$. 
It should be noted here that, even though 
the total rate of events is thus underestimated, 
the rate-weighted integral of $S$ remains conserved, that is
\begin{eqnarray}
\sum_i \lambda_a(S_i) & \leq & \sum_i \lambda(S_i)\\
\label{inequality}
\sum_{i} S_i \lambda_a(S_i) & = & \sum_{i} S_i \lambda(S_i)
\label{conservation}
\end{eqnarray}
Let us now consider events of a given fixed attribute value $S_0$ 
and examine the probability of occurrence of one or more of such 
events. Since the mutually independent discrete events are expected to
follow Poisson statistics,
the probability of, in general, 
$k$ ($\ge 0$) such events occurring within the resolution $\Delta$, 
is given by the function
\begin{equation}
P_{poisson}(k;\lambda(S_0)) = \frac{{[\lambda(S_0)]}^{k} e^{-\lambda(S_0)}}{k!} 
\end{equation}
where $\lambda$ is again the mean rate of events, or more specifically, 
the average number of events expected per $\Delta$. Thus $k_i$ events, 
each with same attribute value $S_i$, occurring within $\Delta$ 
would be unresolved, and hence, they together would be mistaken 
for one event of strength $S_{i}^{a}=k_iS_i$. Thus, even though the true rate 
distribution is non-zero only at $S=S_i$, the {\it apparent} 
probability density distribution (PDD) spreads 
to all non-negative integral multiples of $S_i$ when events are 
viewed with resolution $\Delta$, as 
\begin{equation}
P_{i}^{a}(S_{i}^{a}=k_iS_i) = \frac{{[\lambda(S_i)]}^{k_i} e^{-\lambda(S_i)}}{k_i!} 
\label{single_attr_pdf}
\end{equation}
with an implicit maximum event count of one per $\Delta$.

In general, for an ensemble of mutually independent discrete events 
with a range of the {\it true} attribute values (say, $S_i$, $i=1 \to N$), 
the resultant PDD across the apparent attribute value $S^{a}$ ($=\sum_{i=1}^{N} S_{i}^{a}$)
would be a grand convolution of the {\it apparent} PDDs (as in equation \ref{single_attr_pdf})
corresponding to each of the respective apparent value $S_{i}^{a}$.

\begin{eqnarray}
P^a(S^a=\sum_{i=1}^{N} k_iS_i) & = & \bigotimes_{i=1}^{N} P_{i}^{a}(S_{i}^{a} = k_iS_i) \nonumber \\
                              & = & \bigotimes_{i=1}^{N} \frac{{[\lambda(S_i)]}^{k_i} e^{-\lambda(S_i)}}{k_i!} 
\end{eqnarray}
where $\bigotimes$ denotes convolution product,
and again, $k_i$ is the number of events with {\it true} attribute $S_i$, occurring within $\Delta$.

If ${\widetilde{P}}^{\,a}(f)$ \& ${\widetilde{P}}_{i}^{\,a}(f)$ 
are the Fourier transforms of $P^a(S^a)$ \& $P_{i}^{a}(S_{i}^{a})$ respectively, 
the {\it convolution theorem} would relate them as follows,
\begin{equation}
{\widetilde{P}}^{\,a}(f)= \prod_{i=1}^{N} {\widetilde{P}}_{i}^{\,a}(f) 
\label{conv_thm}
\end{equation}
where $\prod$ denotes a simple product.

The Fourier transforms appearing in the product on the 
right-hand side of the above equation, {\it i.e.} 
${\widetilde{P}}_{i}^{\,a}(f)$, can be obtained in general 
for any $i$, by summing over $k_i$ the Fourier contribution 
from each of the components of $P_{i}^{a}(S_{i}^{a})$ (see equation \ref{single_attr_pdf}), 
evaluated at the discrete values of $S_{i}^{a}=k_iS_i$. Thus,
{
\begin{eqnarray}
{\widetilde{P}}_{i}^{\,a}(f) & = & \sum_{k_i=0}^{\infty} \left ( \frac{{[\lambda(S_i)]}^{k_i} e^{-\lambda(S_i)}}{k_i!}\right ) e^{-j2\pi k_i S_i f} \nonumber \\
 & =  & e^{-\lambda(S_i)} \sum_{k_i=0}^{\infty} \frac{{[\lambda(S_i) e^{-j2\pi S_i f}]}^{k_i}}{k_i!} \nonumber \\
 & =  & e^{-\lambda(S_i)} e^{[\lambda(S_i) e^{-j2\pi S_i f}]} 
\label{single_ft}
\end{eqnarray}
}\\

\begin{figure*}[t!]
\centering
\includegraphics[scale=0.7,angle=-90]{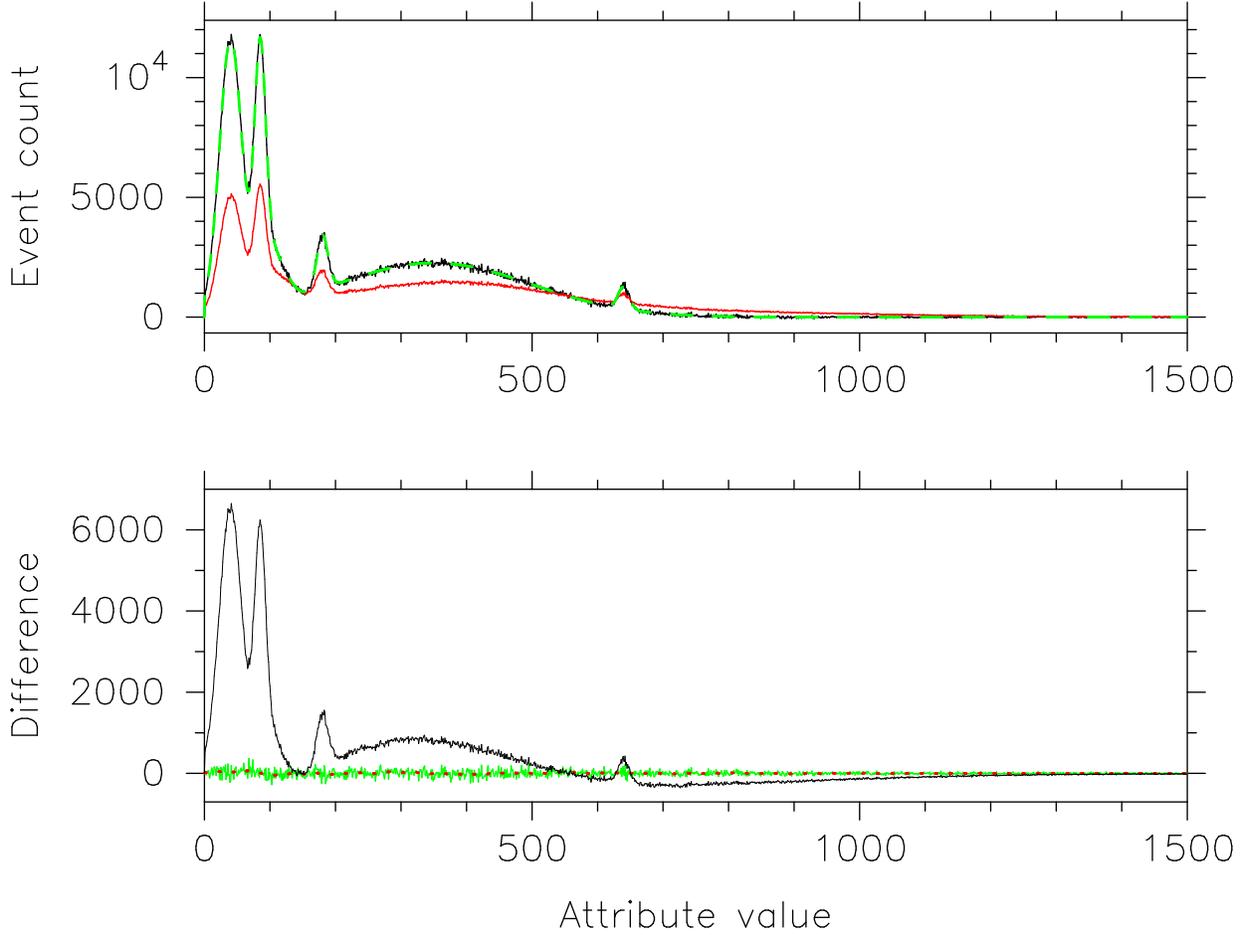}
\caption{Illustration of a distribution affected by pile-up (top panel; red) which is obtained 
through Monte-Carlo simulations, from an assumed model for a ``true" distribution (top panel; green).
The resultant distribution after pile-up correction (top panel; black) 
matches the model ``true" distribution,  within the statistical uncertainties. 
The bottom panel shows the amount 
of estimated correction (bottom panel; black), the difference with the recovered and the model
distributions (bottom panel; green), along with a smoothened version of the latter (bottom panel; red). 
}
\end{figure*}

Substituting this result in equation \ref{conv_thm}, we get 
{
\begin{eqnarray}
{\widetilde{P}}^{\,a}(f) & = & \prod_{i=1}^{N} e^{-\lambda(S_i)} \;\; e^{[\lambda(S_i) \; e^{-j2\pi S_i f}]} \nonumber \\
                 & = & e^{-\sum_{i=1}^{N} \lambda(S_i)} \;\; e^{\sum_{i=1}^{N} \lambda(S_i)\; e^{-j2\pi S_i f}}  \nonumber \\
                 & = & e^{-\sum_{i=1}^{N} \lambda(S_i)} \;\; e^{\widetilde{P}(f)}  
\label{prod_ft}
\end{eqnarray}
}\\
where $\widetilde{P}(f) = \sum_{i=1}^{N} \lambda(S_i)\; e^{-j2\pi S_i f}$, which is the Fourier transform of the
{\it true} distribution. 

By taking natural logarithm of both sides and rearranging, we obtain
{
\begin{eqnarray}
\ln{\left({{\widetilde{P}}^{\,a}(f)}\right)} & = & {\widetilde{P}(f)} \;-\; {\sum_{i=1}^{N} \lambda(S_i)} 
\label{final_truth}
\end{eqnarray}
}

This relation between the Fourier transforms of the {\it true}
and the {\it apparent} distributions of events, 
should enable recovery of the underlying {\it true} rates or counts of events,
as a function of a chosen attribute, from 
the corresponding {\it observed} distribution, often distorted due to pile-up.

The suggested recipe is
\begin{enumerate}
\item  One begins with measurements over a total number of, say, M independent
resolution cells, each of size $\Delta$, providing a record of
discrete events (numbering, say, $N_c$, where $N_c \le M$).
Using such data, the events are sorted and counted
according to the apparent value of their chosen attribute ($S^a$).
The sorted event-count distribution, say $C^{\,a}(S^a)$,
is normalized by M to obtain an apparent probability
density distribution (PDD) of events across $S^a$,
that is $P^{\,a}(S^a) = C^{\,a}(S^a)/M$. The PDD value at $S^a=0$,
if not known a priori or explicitly, can be estimated trivially
and is $\ge 0$, such that the total probability
(including that at $S^a=0$) equals unity.
\item  $P^{\,a}(S^a)$ is Fourier transformed to obtain
a so-called characteristic function, ${\widetilde{P}}^{\,a}(f)$,
but avoiding any normalization by the number of points (N) transformed.
\item  This crucial step involves computing
$\widetilde{X}(f) = \ln{\left({{\widetilde{P}}^{\,a}(f)}\right)}$,
such that if ${{\widetilde{P}}^{\,a}(f)} = a(f) e^{j\phi(f)}$, then
$\widetilde{X}(f) = \ln{\left(a(f)\right)} + j\phi(f)$, where
$j = \sqrt{-1}$.
\item  Inverse Fourier transforming $\widetilde{X}(f)$
(now with usual normalization by N) gives
the {\it true} event-rate distribution $\lambda(S)$ across
the $S$ range, along with a {\it dip} at $S=0$ whose magnitude
is $\sum_{i=1}^{N} \lambda(S_i)$ for $S_i\ne 0$.
\item At $S=0$, one may ignore this (dip) contribution completely,
or compare its magnitude with the integral over the rest of the $S$-range
to assess internal consistency.
The $\lambda(S)$ thus obtained is multiplied by M to get the
estimate of true distribution of event counts $C(S)$,
or further divided by $\Delta$ to get the underlying
event-rate distribution in relevant basic units (such as per unit area
and/or per unit time).
\end{enumerate}

It is important to satisfy the constraint, integral of 
$P^a(S^a)$ being equal to unity, in general, and also to ensure that the 
rate-weighted integral of $S$ is conserved as desired 
(see Equation \ref{conservation}) 
through the above restoration procedure.  
It is easy to show from Equation \ref{final_truth} that
{
\begin{equation}
\left(\frac{d{\widetilde{P}}^{\,a}}{df} \right)_{f=0}  =  {\widetilde{P}}^{\,a}({\scriptstyle{f=0}}) \;\; \left(\frac{d{\widetilde{P}}}{df}\right)_{f=0}
\label{constraint}
\end{equation}
}
\\
where the first derivatives of $\widetilde{P}$ and 
${\widetilde{P}}^{\,a}$ with respect to $f$, when evaluated 
at $f=0$, correspond to respective rate-weighted integrals of $S$, 
or to the first moments of the respective distributions.
In contrast with the above, the true event-rate distribution $\lambda(S)$
is not expected to follow any such constraint on its integral.

Figure 1 illustrates application of our procedure, and its result.
For simplicity, and without loss of generality, all distributions have been 
binned with $S$-interval of unity.
The level of correction effected by the procedure is clearly evident
on the left-side part of the distribution, where this {\it home-coming} of counts is
accompanied by corresponding {\it deportation} out of the right-side region. 
In the present example, the restored count totals to about 1.7 million 
(consistent with the original model distribution), 
of which about 32\% was lost due to pile-up. Although the original
distribution is confined to attribute values $S < 900$, the apparent 
distribution extended well beyond due to pile-up, with about 45000 counts in the range $S > 900$.
We note that the restored distribution is seen also confined to $S < 900$, and
at larger $S$, any deviation now from the expected count (i.e. zero) is found to
be well within statistical uncertainties.

It is important to emphasize that the above procedure for
pile-up correction works equally well also for 
two-sided distributions. Note that when the distribution $P^a(S^a)$
is one-sided (either $S \ge 0$, as in Figure 1, or $S \le 0$), the real and imaginary
parts of ${\widetilde{P}}^{\,a}(f)$ represent a Hilbert pair, and so do
the corresponding parts of $\ln{\left({{\widetilde{P}}^{\,a}(f)}\right)}$,
consistent with the recovered $P(S)$ also being one-sided.
Also, the derived relation (Eq. \ref{final_truth}) provides
a direct way for predicting a piled-up distribution, if the
{\it true} distribution is known. 

\section*{Applications and Discussion}
Pile-up effects have been encountered and discussed in a wide
variety of contexts and measurements over the past several decades,
as apparent from the non-exhaustive list below.
\begin{enumerate}
\item In X-ray astronomy (e.g. the Suzaku \cite{Yam11} and the Chandra \cite{Dav01} missions), 
particularly where CCDs or similar detectors
are employed for measuring (photon) energy spectra, apart from imaging, of 
celestial sources.
\item In high-energy physics experiments\cite{RJ83} (e.g. using the Large 
Hadron Collider at CERN, including the on-going hunt for
Higgs boson\cite{GBrumfiel12}),
particle detectors are equipped with triggers to evaluate 
interaction among high energy particles which are being studied. 
Cosmic ray detectors are also equipped with 
similar triggers which are enabled when cosmic rays of 
sufficiently high energy enter the Cherenkov detectors \cite{Jon04}. 
In both contexts, {\it fake} triggers can occur
due to pile-up of multiple events of lower energy, and on the other hand,
using higher thresholds (to reduce such {\it ghost} or {\it phantom} particles) can 
``miss'' to detect a real particle. 
\item In radiation measurement application and/or 
in Gamma spectroscopy etc., using solid state detectors (e.g. Si(Li); NaI(Tl))\cite{GL99,Mow11,MDF11}
\item In medical clinical diagnostics \cite{APW02,LBL02,Wong01} such as radio-nuclide 
therapy dosimetry imaging, micro-dosimetry of inhaled $\alpha-$emitters 
(e.g. in measurements of specific energy spectra of epithelial
cells of bronchiolar airways) and cardiac first-pass imaging, 
using Gamma Cameras. 

\item In neutrino mass determination, using 
micro-calorimeter to measure the entire spectrum of $^{187}Re$ 
(MARE experiment \cite{Ped08}). 
\end{enumerate}
The prevailing approaches to tackle the pile-up issue are:
a) use as high a spatio-temporal resolution as possible,
b) reduce rate of events, if controllable, or
c) restrict to regions with significantly reduced event rates,
thus ignoring potentially valuable data from regions 
that would be rich in events. The latter two result in 
poor statistics, compromising sensitivity of measurements.
In the absence of any correction procedure so far, 
iterative procedures\cite{GL99,Ped08,Mow11,MDF11}
to seek the underlying ``true" distribution are in use, wherein
Monte-Carlo simulations of events, following assumed models of 
the ``true" distribution and of pile-up, are employed to 
obtain simulated apparent distributions that are compared with 
those measured.

The simple relation derived by us (Eq. \ref{final_truth})
and the correction procedure presented here should find 
ready applicability, in the above mentioned and other
relevant areas, for recovering the underlying {\it true}
distribution of events, even when the observed distributions
have significant distortion due to pile-up.
This, in turn, would enable such measurements with significantly
improved sensitivity, and detection of features or events
otherwise masked by the pile-up distortion.

\end{document}